\documentstyle[epsf]{article}
\begin{document}

\large
\noindent
{\bf Induced Parity Violation in Odd Dimensions}
\normalsize
\vspace{.7in}

\noindent
{\em R. Delbourgo and A. B. Waites}
\vspace{.1in}

\small
\noindent
Physics Department, University of Tasmania,\\
 GPO Box 252C, Hobart, Tas. 7001, Australia.
\vspace{.5in}

\noindent
{\em Abstract}

\noindent
One of the interesting features about field theories in odd dimensions is 
the induction of parity violating terms and well-defined {\em finite} 
topological actions via quantum loops if a fermion mass term is originally
present and conversely. Aspects of this issue are illustrated for 
electrodynamics in 2+1 and 4+1 dimensions. 
\normalsize
\vspace{.3in}

\noindent
{\bf 1. Introduction}

There are a few curiosities associated with field theories in an odd 
number of space-time dimensions. The first is that the overall degree
of divergence of an integral possessing an odd mass scale cannot be taken
at face value since such an integral behaves like the gamma function at 
half-integral argument values; this is most easily seen by considering
dimensional regularization of the typical integral as $D$ tends to an
odd integer (when $r$ below is an integer),
\begin{equation}
 I = -i\int \frac{\Gamma(r)\,\bar{d}^D\!p}{(p^2 - M^2)^r}
   = \frac{(-1)^r\Gamma(r-D/2)}{(4\pi)^{D/2}(M^2)^{r-D/2}}.
\end{equation}

A second noteworthy property is that in odd dimensions, one commonly
encounters couplings which are odd powers of mass. This can be understood
by considering a free-field theory in $D$ space-time dimensions,
\begin{equation}
\int d^D\!x\,[((\partial\phi)^2-\mu^2\phi^2)/2+ \bar{\psi}(i\gamma.\partial
             -m)\psi - F^{\mu\nu}F_{\mu\nu}/4 ] ,
\end{equation}
where typically $\phi$ is a scalar, $\psi$ is a spinor and $F_{\mu\nu}
\equiv \partial_\mu A_\nu - \partial_\nu A_\mu$ is a Maxwell gauge field.
The dimensionlessness of the action (in natural units) specifies the mass 
dimensions of the fields,
$$ [\phi], [A] \sim M^{D/2-1}, \qquad [\psi] \sim M^{(D-1)/2}, $$
whereupon interaction Lagrangians like
\begin{equation}
 \int d^D\!x\,[e\bar{\psi}\gamma.A\psi + f\phi^3 + g\phi\bar{\psi}\psi
               + \lambda\phi^4 + G(\bar{\psi}\psi)^2 + \ldots]
\end{equation}
will have coupling constants with prescribed dimensions,
\begin{equation}
 [e],[g] \sim M^{2-D/2}, \quad [f] \sim M^{3-D/2}, 
 \quad [\lambda] \sim M^{4-D}, \quad [G] \sim M^{2-D}.
\end{equation}
One then observes that in odd dimensions the couplings $e,f,g$ have odd
$\sqrt{M}$ scales. That does not matter much for electrodynamics
since we meet powers of $e^2$ in the perturbation expansion but for $f,g$ 
we can potentially encounter single powers of the coupling. In particular 
for three dimensions,
\begin{equation}
 [f] \sim M^{3/2}, \qquad [e],[g] \sim M^{1/2}, \qquad [\lambda]\sim M,
 \qquad [G] \sim M^{-1},
\end{equation}
telling us that electrodynamics, chromodynamics, $\lambda\phi^4$ and Yukawa
interactions become {\em super}-renormalizable, while Fermi interactions 
remain unrenormalizable. Combined with property one, this has the effect
of eliminating certain ultraviolet infinities in theories such as $\lambda
\phi^4$; thus the tadpole graphs of order $\lambda$ and higher are
perfectly finite and the {\em only} infinity in that model is the self-mass
of $\phi$ due to the three-$\phi$ intermediate state. For electrodynamics 
in 2+1 dimensions, the situation is even better---no infinities at 
all.

A third peculiar aspect of odd $D$ dimensions stems from the algebra of
the Dirac $\gamma$-matrices 
$$ \{\gamma_\mu,\gamma_\nu\} = 2\eta_{\mu\nu}. $$
When $D$ is even it is well-known that the `$\gamma$' have size
$2^{D/2}\times 2^{D/2}$ and there exists a `$\gamma_5$' matrix which is 
the product of all the different $D~\gamma$'s and which anticommutes with 
each $\gamma_\mu$; one can always arrange it to have square -1, like all 
the space-like $\gamma$ (in our metric +,-,-,..). It is not so well-known
that in one higher dimension, the size of the $\gamma$'s remains the 
same---all that happens is that the `$\gamma_5$' matrix becomes the last 
element of the $D$ Dirac matrices.

For instance, in three dimensions one can take the two-dimensional Pauli
$\gamma_0 = \sigma_3, \gamma_1 = i\sigma_1 $ and simply append
$\gamma_2 = `\gamma_5{\mbox '} = i\gamma_0\gamma_1 = -i\sigma_2$ to complete
the set, without altering the size of the representation. At the same time
it should be noticed that one can get a non-zero trace from the product of 
{\em three} gamma-matrices, viz. Tr$[\gamma_\rho\gamma_\sigma\gamma_\tau]=2i
\epsilon_{\rho\sigma\tau}$. 
Similarly, in five dimensions one can take the usual four-dimensional ones
and just append $\gamma_5 \equiv \gamma_0\gamma_1\gamma_2\gamma_3$ as the 
fifth component; here as well the product of the full five $\gamma$'s gives 
a non-vanishing trace: Tr$[\gamma_\mu\gamma_\nu\gamma_\rho\gamma_\sigma
\gamma_\tau]= -4\epsilon_{\mu\nu\rho\sigma\tau}$. The lesson is that when 
$D$ is odd one should be careful before discarding traces of odd 
monomials of gamma-matrices, if there are sufficiently many `$\gamma$'s 
since at least
$${\rm Tr}[\gamma_{\mu_1}\gamma_{\mu_2}...\gamma_{\mu_D}]=(2i)^{(D-1)/2}
              \epsilon_{\mu_1\mu_2\cdots\mu_D} .$$
Another property worth remembering in odd dimensions is that if one 
constructs suitably normalized antisymmetric products of $r$ matrices, 
$\gamma_{[\mu_1\mu_2\cdots\mu_r]} $ (the total set of these from $r=0$ to 
$r=D$ generates a complete set into which any $2^{[D/2]}\times2^{[D/2]}$ 
matrix can be expanded) then there exists the relation,
$$\gamma_{[\mu_1\mu_2\cdots\mu_r]}=i^{(D-1)/2}\epsilon_{\mu_1\mu_2\cdots
        \mu_D}\gamma^{[\mu_{r+1}\mu_{r+2}\cdots\mu_D]}/(D-r)! $$
This often helps in simplifying products of matrices.

Turning to discrete operations, a charge conjugation operator ${\cal C}$ 
with the transposition property
\begin{equation}
 {\cal C}\gamma_{[\mu_1\mu_2\cdots\mu_r]}{\cal C}^{-1} = (-1)^{[(r+1)/2]}
         (\gamma_{[\mu_1\mu_2\cdots\mu_r]})^T
\end{equation}
always exists in even dimensions, but cannot be defined at the odd values
$D=5,9,13..$. This is intimately tied with the existence of topological 
terms in the action for the pure gauge field, as we shall see. As for
parity ${\cal P}$, it corresponds to an inversion of all the {\em
spatial} coordinates for even $D$, since that is an improper transformation.
However when $D$ is odd it should be regarded as a reflection of all the 
space coordinates {\em excepting the very last one}, $x_{D-1}$, in order
to ensure that the determinant of the transformation remains negative. 
It is straightforward to verify that this corresponds to the unitary 
change,
\begin{eqnarray}
 {\cal P}\psi(x_0,x_1,..,x_{D-2},x_{D-1}){\cal P}^{-1}\!\!\!&=&\!\!\!
 -i\eta\gamma_0\gamma_{D-1} \psi(x_0,-x_1,..,-x_{D-2},x_{D-1})\nonumber \\ 
 \!\!\!&=&\!\!\!\eta\gamma_1\cdots\gamma_{D-2}\psi(x_0,-x_1,..,
 -x_{D-2},x_{D-1}),
\end{eqnarray}
where $\eta$ is the intrinsic parity of the fermion field. In that regard,
one can just as easily check that a mass term like $m\bar{\psi}\psi$ in
the action is {\em not} invariant under parity for $D$ odd. 

This potential to induce other parity-violating interactions in the theory 
forms the subject of this paper. In Section 2 we shall examine the 
induction of a Chern-Simons term in 2+1 electrodynamics from a fermion
mass; the converse process is considered in an Appendix. In Section 3 we
generalize this to 4+1 QED and to higher $D$, with the converse effect
also treated in the Appendix. An explanation of why this is a pure 
one-loop effect is also provided. Finally we discuss in Section 4 what
happens when electrodynamics is purely topological (no free $F^2$ term in 
the Lagrangian) as this represents a system quite different from what we
are accustomed to.

\vspace{.2in}

\noindent
{\bf 2. 3D Electrodynamics}

Turning to QED in 2+1 dimensions, we are blessed with a coupling with
positive mass dimension $[e^2] \sim M$, so we anticipate a finite number
of ultraviolet singularities. But in fact {\em none exists} thanks to
gauge invariance. The standard diagrams for photon polarization, electron
self-energy and vertex corrections are all perfectly finite---barring 
infrared problems, which actually correct themselves non-perturbatively 
through the dressing of the photon line, as first demonstrated by Jackiw 
and Templeton (1981), although some doubts about the loop expansion have 
been expressed by Pisarski and Rao (1985). 

Straightforward evaluation of the graphs produces the one-loop results,

\begin{eqnarray}
 \Pi_{\mu\nu}(k)&=& (k_\mu k_\nu - k^2\eta_{\mu\nu})\frac{e^2}{2\pi}
   \int \frac{\alpha(1-\alpha) d\alpha}{\sqrt{m^2-k^2\alpha(1-\alpha)}}
   \nonumber \\
   & & + im\epsilon_{\lambda\mu\nu}k^\lambda \frac{e^2}{4\pi}
         \int \frac{d\alpha}{\sqrt{m^2-k^2\alpha(1-\alpha)}}
\end{eqnarray}
\begin{equation}
 S(p) = \langle T(\bar{\psi}(p)\psi(0))\rangle =
       \frac{e^2}{16\pi}\int \frac{dw}{w(\gamma.p-w)}
       \left[\frac{\xi}{w} - \frac{4m}{(w-m)^2}\right],
\end{equation}
where $\xi$ is a parameter that fixes a Lorentz-covariant gauge ($\xi=0$ is
the Landau gauge). The vertex corrections are obviously finite too, because
`$Z_1 = Z_2$'. Higher powers of $e^2$ only serve to make the diagrams more
convergent than they already are in lowest order, since the series 
expansion of a physical amplitude will take the form,
$$ T(k)=T_0(k)\left[1 + c_1\frac{e^2}{m}+c_2\frac{e^4}{m^2}+\ldots\right];
 \quad c_i=f(k/m), $$
where $k$ signify the external momentum variables. Broadhurst, Fleischer 
and Tarasov (1993) have calculated some of these coefficients for any $D$.

An important aspect of (8) is the induced parity-violating Chern-Simons 
interaction, $\epsilon_{\lambda\mu\nu}A^\lambda F^{\mu\nu}$. It is not
surprising that it should have sprouted, since we started with a fermion 
mass $m\bar{\psi}\psi$ term which is intrinsically ${\cal P}$-violating in 
odd dimensions; but the value of that induced photon term is finite and 
disappears as $m \rightarrow 0$. Note however that in the infrared limit 
it reduces to $ie^2\epsilon_{\lambda\mu\nu}k^\lambda/4\pi$ {\sl provided 
that} $m\neq 0$. 

There has been some argument in the literature that this Chern-Simons term 
may produce anomalies in some processes because it contains an $\epsilon$ 
tensor specific to three dimensions (in much the same way that the axial 
anomaly is connected with the $\epsilon$ tensor in four dimensions). 
This cannot be true because vacuum polarization to order $e^2$ is perfectly
sensible and {\em finite} when evaluated by any reasonable regularization. 
Anomalies only arise when a {\em divergence} multiplies an apparent 
{\em zero}: in the Pauli-Villars method, when a mass 
regulator contribution $M^2$ multiplies an integral of order $1/M^2$; in a 
dimensional context when a pole $1/(D-3)$ multiplies an evanescent zero 
$(D-3)$; but as we have already seen the photon self-energy diagram 
contains no such singularity (Delbourgo and Waites, 1993). Thus an anomaly 
is indeed absent.
\vspace{.2in}

\noindent
{\bf 3. Electrodynamics in Higher Dimensions}

A subject of debate has been whether the induced Chern-Simons term 
suffers from higher order loop corrections. There is a simple proof
presented later which says this cannot be, but before describing
it let us exhibit the induced term for any odd $D$ dimension as it is a
`clean' result. We begin as before with massive fermion QED. For
arbitrary $D$ the induced topological term takes the form of an $n$-point
function,
\begin{equation}
 C\epsilon_{\mu_1\mu_2\cdots\mu_D}A^{\mu_1}F^{\mu_2\mu_3}\cdots
                                 F^{\mu_{D-1}\mu_D}; \qquad n=(1+D)/2.
\end{equation}
Notice that this conforms perfectly with charge conjugation: when $D=3$
and ${\cal C}$ is conserved, the topological term involves an even number
$n=2$ of photons; when $D=5$ and $[e^2] \sim M^{-1}$, we encounter three 
photon lines but then ${\cal C}$ is no longer valid; when $D=7$, 
${\cal C}$-invariance becomes operative again and the number of photon 
lines is $n=4$, and so on.

The result of the one-loop contribution to the topological term has 
already been quoted in Eq.(8). Looking at the next odd dimension, $D=5$,
the relevant one-loop graphs are in Figure 1, leading to the
induced vertex,
$$\Gamma_{\lambda\mu\nu}(k,k')=-2ie^3\!\!\!\int\!\!\bar{d}^5\!p\frac
{{\rm Tr}[\gamma_\nu(\gamma.p+m)\gamma_\mu(\gamma.(p+k)+m)\gamma_\lambda
         (\gamma.(p-k')+m)]} {(p^2-m^2)((p+k)^2+m^2)((p-k')^2-m^2)}. $$
Introducing Feynman parameters in the usual way to combine denominators and
picking out the term with five gamma-matrices in the trace, we end up with
\begin{eqnarray}
 \Gamma_{\lambda\mu\nu}(k,k')\!\!&=&\!\!\!-16ie^3m\!\!\int\!\!\bar{d}^5\!p
              \frac{d\alpha d\beta d\gamma\,\delta(1-\alpha-\beta-\gamma)
                    \epsilon_{\lambda\mu\nu\rho\sigma}k^\rho k'^\sigma}
{[p^2-m^2+k^2\alpha\beta+k'^2\gamma\alpha+(k+k')^2\beta\gamma]^3}\nonumber\\
\!\!&=&\!\!\!-\frac{e^3m\epsilon_{\lambda\mu\nu\rho\sigma}k^\rho k'^\sigma}
 {8\pi^2}\!\!\int_0^1\!\!\!\!\!\frac{d\alpha d\beta d\gamma\,
 \delta(1-\alpha-\beta-\gamma)}
     {\sqrt{m^2-k^2\alpha\beta-k'^2\gamma\alpha-(k+k')^2\beta\gamma}}
\end{eqnarray}

\begin{figure}
\hspace*{2cm}\epsfxsize=8cm \epsfbox{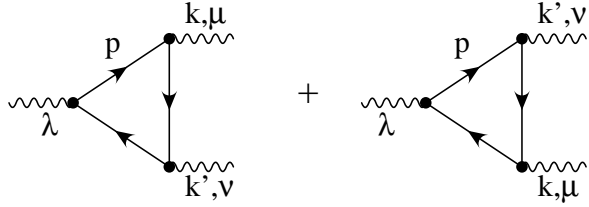}
 \caption{One-loop induction of a Chern-Simons amplitude in 5 dimensions.}
\end{figure}

One can regard this amplitude as the five-dimensional description of the
process $\pi^0 \rightarrow 2\gamma$, because one of the indices (4) of the
Levi-Civita tensor just corresponds to the standard pseudoscalar and the
residual four indices (0 to 3) are the normal 4-vector ones. Just as with 
2+1 QED, we see that the induced term in 4+1 QED vanishes with the fermion 
mass $m$. [Contrariwise one can check that if $m=0$, but a term (10) is 
present from the word go, then a fermion mass term amongst other 
parity-violating ones will arise. See the Appendix.]

We are now in a position to quote the topological vertex induced for 
arbitrary odd $D$ by the fermion mass term. Introduce $n=(D+1)/2\quad$
Feynman parameters $\alpha_i; i=1 \ldots n$ for each internal line 
(Figure~2). Call the momentum flowing across each possible cutting of 
two lines $k_{ij}$ if those lines have parameters $\alpha_i,\alpha_j$. 
The calculation then produces the result,
\begin{eqnarray}
 \Gamma_{\mu_1\mu_2\cdots\mu_n}(k)&=&-\frac{me^ni^{n-1}}{2(2\pi)^{n-1}}
 \epsilon_{\mu_1\mu_2\cdots\mu_D}k_1^{\mu_2}k_2^{\mu_4}\ldots k_n^{\mu_D}.
  \nonumber \\
  & & .\int_0^1\Pi_{k=1}^n d\alpha_k\frac{\delta(1-\sum_k\alpha_k)}
          {\sqrt{m^2-\sum_{i<j=1}^n k_{ij}^2\alpha_i\alpha_j}}.
\end{eqnarray}

\begin{figure}
\hspace*{2.5cm}\epsfxsize=7cm \epsfbox{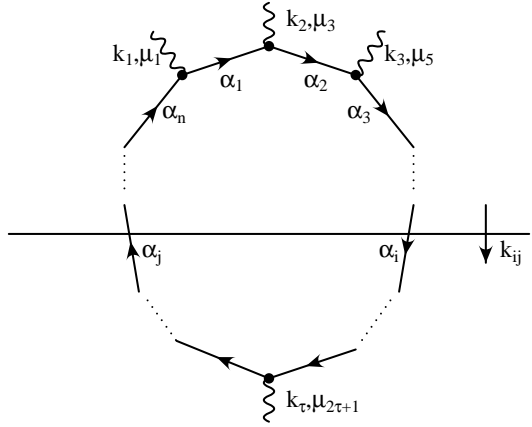}
 \caption{One-loop induction of a Chern-Simons term in $D$ dimensions.}
\end{figure}

\noindent
One may readily check that this collapses to the results (8) and (11) for
$D=3$ and $D=5$ respectively. It corresponds to the Chern-Simons term (10)
where $C = e^n/2n!(4\pi)^{n-1}$ if one goes to the soft photon limit,
always assuming $m \neq 0$. 

\begin{figure}
\hspace*{3cm}\epsfxsize=6cm \epsfbox{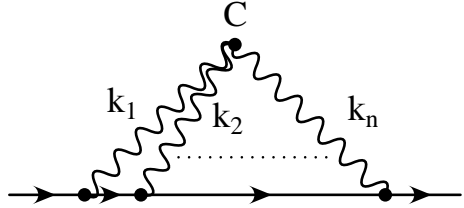}
\caption{Induction of a fermion mass term through a topological 
         interaction.}
\end{figure}

To finish off this section let us explain why this one-loop answer (12) is 
all there is. In 3 dimensions, the Lagrangian 
$\epsilon_{\lambda\mu\nu}A^\lambda F^{\mu\nu}$
will change by a pure divergence under the gauge transformation, $\delta A
\rightarrow \partial\chi$, so the action remains invariant for all normal
field configurations that vanish at $\infty$. However a fourth order
interaction like
$$\epsilon_{\lambda\mu\nu}A^\lambda F^{\mu\nu}F_{\rho\sigma}F^{\rho\sigma}$$
will {\em not} be invariant under the gauge change; thus it is not
permitted. More generally, in odd $D$ dimensions the interaction
\begin{equation}
 \epsilon_{\mu_1\mu_2\cdots\mu_D}A^{\mu_1}F^{\mu_2\mu_3}\ldots
           F^{\mu_{D-1}\mu_D}(F_{\rho\sigma}F^{\rho\sigma})^N;
           \qquad N \geq 1,
\end{equation}
and ones like it, are forbidden by gauge invariance and thus cannot be 
produced. On the other hand a two-loop contribution to the fundamental
topological term can be regarded as an integration of (13), with $N=1$, 
over one of the photon momenta. Since we have just concluded that (13) must
be absent, we deduce that the induced topological term (10) cannot receive 
any two-loop (or higher-loop) quantum corrections. In that respect, it is a
pristine result similar to the Adler-Bardeen theorem for the axial anomaly;
nevertheless it is only of academic interest in as much as QED becomes
unrenormalizable (c.f. the dimensions of $e^2$) when $D \geq 5$, unless the
space-time is compact, e.g. in some Kaluza-Klein geometries.
\vspace{.2in}

\noindent
{\bf 4. Topological QED}
   
So far we have considered models where the initial Lagrangian contains
the normal free gauge kinetic energy $F_{\mu\nu}F^{\mu\nu}$ term and 
seen what transpires as a result of parity violation primarily through the
fermion field. Now we shall consider what happens when the initial 
Lagrangian has no gauge field kinetic energy but starts off life instead 
with a Chern-Simons piece such as (10). In 2+1 dimensions, this still means
a bilinear term in the gauge field capable of launching a propagator,
\begin{equation}
 D_{\mu\nu} = [\frac{i\epsilon_{\mu\nu\lambda}k^\lambda}{\mu k^2}
               -\xi \frac{k_\mu k_\nu}{k^4}],   
\end{equation}
where we have taken account of gauge-fixing, with parameter $\xi$. (The very 
same expression can be obtained by adding a conventional kinetic term 
$-Z F_{\mu\nu}F^{\mu\nu}/4$ and taking the limit $Z \rightarrow 0$.) 
Evaluating the fermion self-energy now yields,
\begin{equation}
 \Sigma(p) = e^2\int \bar{d}^3k\; 
      \frac{\gamma_\mu(\gamma.p-\gamma.k)\gamma_\nu}{(p-k)^2}D^{\mu\nu}(k)
          = -e^2\left[\frac{\xi\gamma.p}{16\sqrt{-p^2}} + 
            \frac{\sqrt{-p^2}}{8\mu}\right]                                  
\end{equation}
containing a mass-like term where none previously existed. In the same vein 
we may compute the vacuum polarization correction to (14) and arrive at
\begin{equation}
 \Pi_{\mu\nu}(k) = ie^2 {\rm Tr}\int \bar{d}^3p\; \frac{\gamma_\mu\gamma.p
                   \gamma_\nu\gamma.(p - k)}{p^2(p-k)^2}
                 = (-k^2\eta_{\mu\nu}+k_\mu k_\nu)\frac{e^2}{8\sqrt{-k^2}},
\end{equation}
which has the effect of leaving $D(k) \sim 1/k$. In higher orders of 
perturbation theory we may expect that
$$\Sigma(p) = \gamma.p\; f(\frac{e^2}{\sqrt{-p^2}},\frac{e^2}{\mu}) +
              \sqrt{-p^2}\;g(\frac{e^2}{\sqrt{-p^2}},\frac{e^2}{\mu});$$
               
$$ \Pi_{\mu\nu} = (-k^2\eta_{\mu\nu} + k_\mu k_\nu)\;
                  \pi(\frac{e^2}{\sqrt{-k^2}},\frac{e^2}{\mu}), $$
where $f,g$ and $\pi$ are scalar functions of their arguments. It is
fascinating to speculate on the full form of those functions by
applying some non-perturbative method of solution.                        

The situation is radically different in 4+1 dimensions since the 
Chern-Simons term is {\em trilinear} in the gauge field and alone cannot 
engender a propagator. Rather one must resort to quantum corrections to get
something of that ilk; the vacuum polarization graph (initially from 
massless fermions) produces a hard quantum loop contribution:
\begin{equation}
D_{\mu\nu} = (-\frac{\eta_{\mu\nu}}{k^2} + \frac{k_\mu k_\nu}{k^4})
                \frac{512\pi}{3e^2\sqrt{-k^2}} -\xi\frac{k_\mu k_\nu}{k^4}.
\end{equation}                     
Taken with the trilinear gauge interaction this can produce a vacuum 
polarization effect from the gauge field itself, namely
$$\Pi_{\mu\nu} = i(\frac{512\pi C}{3e^2})^2 \int \frac{\bar{d}^5k'}
                   {(k'^2(k-k')^2)^{3/2}}
                  \epsilon_{\mu\rho\sigma\alpha\beta}k^\alpha k'^\beta
                {\epsilon_\nu}^{\rho\sigma\gamma\delta}k_\gamma k'_\delta$$
or
$$\Pi_{\mu\nu}  = (\eta_{\mu\nu}k^2 - k_\mu k_\nu)\left(\frac{512C}{3e^2}
\right)^2\frac{\sqrt{-k^2}}{12}. $$
Interestingly, (17) does not give birth to a mass-like fermion self-energy 
at one loop level---five gamma matrices are needed to obtain that. This 
means we have to consider two loop effects, either to order $e^4$ or to 
first order in the Chern-Simons coupling $C$, as sketched in Figure 3. 
Quite generally we may anticipate in 4+1 dimensions that
$$\Sigma(p) = \gamma.p\;f(e^2\sqrt{-p^2},C/e^3) +
              \sqrt{-p^2}\;g(e^2\sqrt{-p^2},C/e^3);$$
               
$$\Pi_{\mu\nu} = (-k^2\eta_{\mu\nu} + k_\mu k_\nu)\;
                       \pi(e^2\sqrt{-k^2},C/e^3). $$
However we must be on guard that higher order contributions in $e^2$ and 
$C$ are very likely unrenormalizable now and possibly of academic interest.
Still, our discussion does indicate the nature of such parity-violating 
contributions in these models and how they spring from just one source.

If we could trust some non-perturbative method for resumming
the Feynman diagrams then we might be able to estimate the quantum effects
associated with the dimensional couplings $e^2$ and $C$. The same of course
applies with even greater force to pure Chern-Simons theory in higher 
odd-dimensions.
   
\vspace{.3in}

\noindent
{\bf Acknowledgements}

We thank Dirk Kreimer for discussions. The ARC, under Grant A69231484, 
have kindly supported this research.
\vspace{.3in}

\noindent   
{\bf References}
\small

\noindent   
Jackiw, R. and Templeton, S. (1981). {\em Phys. Rev.} D {\bf 23}, 2291.

\noindent
Pisarski, R. and Rao, S. (1985).  {\em Phys. Rev.} D {\bf 32}, 2081.

\noindent
Broadhurst, D., Fleischer, C. and Tarasov, V. (1993). {\em Z. Phys.}
 C {\bf 60}, 287.

\noindent
Delbourgo, R. and Waites, A. B. (1993) {\em Phys. Lett.} B {\bf 300}, 241. 
   
\newpage
\normalsize
\noindent
{\bf Appendix}

Here we shall examine the converse of sections 2 and 3, in as much as we deal
with massless electrodynamics ($A$ and $\psi$) but introduce the parity 
violation through a primary Chern-Simons term, not a fermion mass. Our 
treatment is to be contrasted with section 4, where a kinetic term for the 
photon was absent. In the present circumstances,
$${\cal L} = \bar{\psi}\gamma.(i\partial - eA)\psi - F_{\mu\nu}F^{\mu\nu}/4
             +C\epsilon_{\mu_1\mu_2\cdots\mu_D}A^{\mu_1}F^{\mu_2\mu_3}\cdots
                                 F^{\mu_{D-1}\mu_D},$$ 
we can be certain the gauge field will propagate at the bare level in any
dimension $D$. It so happens that when $D=3$ the Chern-Simons piece is also
bilinear and can be incorporated with the standard $F^2$ term to give
the initial two-point function,
$$D_{\mu\nu} = \frac{-\eta_{\mu\nu} + k_\mu k_\nu/k^2}{k^2-\mu^2}
             + i\frac{\mu\epsilon_{\mu\nu\lambda}k^\lambda}{k^2(k^2-\mu^2)}
             - \xi\frac{k_\mu k_\nu}{k^4}. $$
Parity violating terms then arise through quantum corrections in other Green
functions.

Probably the most significant of these is in the fermion self-energy,
\begin{eqnarray*}
 \Sigma(p)&=&\frac{e^2\gamma.p}{16\pi p^2}\left[\frac{p^4-\mu^4}
              {\mu^2p}\ln\;(\frac{\mu+p}{\mu-p}) - \frac{2(p^2-\mu^2)}{\mu}
              +\frac{\pi p^2\sqrt{-p^2}}{\mu^2}-\frac{3}{2}\pi\xi\sqrt{-p^2}
              \right]  \\
          & &+\frac{e^2}{8\mu\pi}\left[\frac{p^2-\mu^2}{p}\ln\;(\frac{\mu+p}
               {\mu-p}) - 2\mu + \pi\sqrt{-p^2} \right]          
\end{eqnarray*}   
We should notice that in the limit of small $\mu$, the expression reduces to
$$\Sigma(p) = \frac{e^2\gamma.p}{4\pi p^2}[\frac{\mu}{2} - 
               \frac{3\pi\xi\sqrt{-p^2}}{8}]+\frac{e^2\mu}{8\sqrt{-p^2}}.$$
and could have been directly evaluated by regarding $\epsilon AF$ as an
interaction, rather than combining it with the bare photon propagator as 
above. (It will disappear of course when $\mu\rightarrow 0$ in the Landau 
gauge.)

Indeed that is the only sensible treatment for higher dimensions $D$ since the
Chern-Simons term is no longer bilinear. For $D=5$ to first order in 
$C\epsilon AFF$, one engenders the mass term (and no kinetic term)
\begin{eqnarray*}
 \Sigma(p) &=& -ie^3 C\int \bar{d}^Dk\;\bar{d}^Dk'\frac
               {\epsilon_{\mu\nu\lambda\alpha\beta}k^\alpha k'^\beta
                \gamma^\mu\gamma.(p-k-k')\gamma^\nu\gamma.(p-k)\gamma^\lambda}
               {k^2 k'^2 (k+k')^2 (p-k-k')^2 (p-k)^2} \\
            &=& \frac{3\Gamma(3-D)p^4e^3C}{(16)^3\pi^4};
\end{eqnarray*}                
unfortunately this is divergent as $D \rightarrow 5$, which is not too
surprising. There is likewise a 2-loop contribution of the same type to the
photon self-energy, but this cannot add a parity violating part to $\Pi$
because such a term would violate gauge invariance for $D=5$ as we have 
already explained in section 3.

\end{document}